# Symmetry-guided data-driven discovery of native quantum defects in two-dimensional materials

Jeng-Yuan Tsai[1], Weiyi Gong,[1] Qimin Yan[1]*

[1]Department of Physics, Northeastern University, Boston, MA 02115, USA

* Correspondence and requests for materials should be addressed to Q.Y. (q.yan@northeastern.edu).

Abstract

Drawing on their atomically thin structure, two-dimensional (2D) materials present a groundbreaking avenue for the precision fabrication and systematic manipulation of quantum defects. Through a method grounded in site-symmetry principles, we devise an comprehensive workflow to pinpoint potential native quantum defects across the entire spectrum of known binary 2D materials. Leveraging both symmetry principles and data-driven approaches markedly enhances the identification of spin defects exhibiting triplet ground states. This advancement is pivotal in discovering NV-like quantum defects in 2D materials, which are instrumental in facilitating a set of quantum functionalities. For discerning the multifaceted functionalities of these quantum defect candidates, their magneto-optical properties are comprehensively estimated using high-throughput computations. Our findings underscore that antisite defects in diverse hosts emerge as prospective quantum defects of significance. Crucially, based on our research, we advocate that the 16 antisites present in post-transition metal monochalcogenides (PTMCs) stand out as a prominent 2D-material-based quantum defect platforms, by their precise defect levels, optimal magneto-optical attributes, and the readily accessible nature of their host materials. This work substantially broadens the repertoire of quantum defects within the 2D material landscape, presenting profound implications for the advancement of quantum information science and technologies.

**Introduction**

At the microscopic level, quantum mechanics establishes the predominant laws of physics, presenting a distinct approach wherein physical phenomena are described through probabilistic representations.[1] The inherent superposition of quantum states introduces the intriguing idea of processing information simultaneously and provides the potential for developing quantum computational devices with exponential superiority over their conventional counterparts. Traditionally perceived as detrimental in solid-state materials, point defects can yield a collection of stable defect states within the band gap. These states, mirroring the localized molecular orbitals within individual molecules, form an atom-like system embedded in a material. When these localized defect states are synergized with methodologies for initialization, manipulation, and detection, they pave the way for an array of quantum applications, including quantum computing, quantum communication, and quantum sensing.[2-5]

One of the well-known quantum defects is the NV center (nitrogen-vacancy) in diamond. Wide bandgap of diamond renders well-defined in-gap defect states and such electronic spins can be initialized, manipulated, and read out at room temperature.[6] The interplay between electronic and nuclear spins and the entanglement of electronic spins among NV centers are performed as quantum operations for error correction and teleport.[7,8] Inspired by NV center, exploration of defects with improved properties and new functionalities has grown rapidly,[9] including defects in SiC,[10-12] Si-V center in diamond,[13-15] and rare-earth ions in oxide crystals.[16] Other defects have been suggested as promising qubit candidates in SiC[17,18] and AlN.[19]

Recent Quantum defects in 2D materials have unique advantages of realizing quantum computing and information processing as it is atomically thin. The intrinsically planar structures impose more controllable creation and manipulation of quantum defects and enhancement of scalability. Generating defects in 2D materials can be achieved by various existing approaches.[20] In addition, atomic-level scanning probe techniques[21] can be used to characterize and manipulate defects. Defects in h-BN such as carbon-vacancy complex $C_B$-$V_N$[22,23] and charged boron vacancy $V_B^{-}$[24] have been proposed as spin qubits. Antisite defects have been identified as a promising spin qubit in group-VI transition metal dichalcogenides (TMDs) as another well-known class of 2D materials.[25]

Beyond spin qubits, optical transitions generated by in-gap defect states and defect-bound excitons have been shown as single-photon emitters (SPEs) for quantum telecommunication applications. While NV center with its low phonon-side band has been recognized as a promising SPE[4], the difficulty of integrating diamond into devices motivates further exploration of novel quantum defect candidates including those in 2D materials. The wide variety of defects in h-BN has been proposed as the sources of wide-spanning emission lines, including vacancies[26-28], substitutions,[29-31] and complexes.[23,32-37] SPEs from TMDs such as oxygen-related defects[38] in $WSe_2$ have also been observed, and the essence of semiconductors of TMDs indicates the emissions consist of defect-bound excitons.[39,40]

As an exciting and emerging research field, the progress beyond the initial progress of experimental demonstration of quantum defects in 2D materials such as h-BN and TMDs calls for the accelerated discovery and rational design of novel quantum defects in 2D compounds for quantum computing and information process.

Here, we show that by imposing symmetry-based strategy and performing high-throughput techniques, the ratio of triplet defects and total defects is around 64%, implying symmetry-based strategy well enhances the efficiency of searching for triplet defects. Among this set of triplet defects, 45 antisite defects out of 41 host materials, including post-transition metal monochalcogenides (PTMCs), transition metal dichalcogenides (TMDs), and transition metal halogenides are screened out as the potential candidates for quantum applications. The criteria are based on qubit operation principles, including well-isolated in-gap defect levels, operational radiative transitions and sublevel splittings of triplet ground state, and allowed intersystem-crossings, which are implemented by the explicit calculations of dipole transition rate, zero-phono line (ZPL), and zero-field splitting (ZFS) based on dipolar interactions.

The dataset provides insights into the relationships between observables such as ZFS and ZPL, and parameters such as atomic radius, spatial localization of levels, and crystal-field splitting. Our analysis reveals that the energy difference between spin states is correlated with the exchange and on-site interactions, indicating that these interactions are crucial for stabilizing a spin-triplet defect.

This finding aligns with the symmetry-based strategy. The qubit operation principles for candidates imposing defect symmetries of $C_{3v}$ and $D_{2d}$ are also shown. Anion antisites and cation antisites in group-III PTMCs are selected as examples of qubit candidates mainly due to PTMCs being novel 2-dimensional materials.[41-45]

**Results**

**Symmetry-enabled strategy of searching for defects with triplet ground states**

Motivated by the insight gained from the identification of antisite defects in TMDs, the symmetry-enabled strategies can greatly accelerate the search for defects with triplet ground states.[25] Considering a three-level system, if the two lower levels are separated by a small energy splitting and occupied by two spin-parallel electrons, a triplet state is more favorable than a singlet state due to the exchange interaction between spin-parallel electrons. The exchange interaction compensates the increase of total energy as an electron occupies a higher level. In a special case, if the two lower levels are degenerate, the triplet ground state is created based on the Hund's rule. Therefore, creating a high-dimensional and/or degenerate defect-level space is a feasible design route for quantum defects with triplet ground states. Whether a given material hosts local sites to create such defects is embraced by symmetry information through the irreducible representations (IRs) of the corresponding crystalline point groups. A direct sum of IRs can decompose a completely reducible representation. In this way, a state space formed by defect levels can be decomposed by IRs. Since IRs have various dimensions/degeneracies, we focus on the point groups with at least one IR of which the dimension is larger than 1. Accordingly, with the space group of the host material, if a crystalline point group has IRs that satisfy the above prerequisite, it is plausible that some defects in the host material have triplet ground states and are worthwhile for further investigation. Note that the local symmetries at any crystal sites of materials must be subgroups of the crystalline point group. To construct defect structures, we then identify promising sites by their local symmetries that have at least one high-dimensional IR. Once these sites are identified, we create simulation systems with defects at the corresponding sites in the host materials. In this work, we consider only intrinsic defects, including antisites and vacancies.

Another important consideration is the charge state because defect levels occupied by a different number of electrons show different spin states. For instance, to host a triplet ground state, one anticipates that the total number of electrons must be even. If it is even, we consider only the neutral charge state; otherwise, we calculate three charge states 1-, 0, and 1, which an even number of electrons. As defect types and charge states are determined, we perform first-principles calculations based on density-functional theory (Figure 1a).

**Data-driven quantum defect discovery based on symmetry-enabled strategy**

We begin the data-driven defect discovery efforts by data mining potential 2D host materials from the C2DB database[46]. The initial criteria are that the host materials have nonzero band gaps based on HSE06 functional, no net magnetic moments, and binary constituents. We then perform first-principles calculations based on the SCAN meta-GGA functional to obtain their band structures. The number of calculated host materials is 523. We select those with band gaps greater than or equal to 1 eV as candidates, and 310 host materials are screened out. We then implement the symmetry-enabled strategy to create defect structures and perform high-throughput computations. The number of calculated defect systems with various charge states is 1144. Among them, there are 627 unique defect systems built upon 235 host materials.

A striking observation of the high-throughput computational results is that 471 defects with various charge states are identified as their triplet ground states. Among them, there are 403 unique defect systems in 199 host materials (see Supplementary Table 1). We further evaluate the efficiency of our symmetry-based discovery strategy by considering two ratios, (1) the number of triplet defects over the number of calculated defects, i.e., 471/1144 ∼ 41%; (2) the number of unique triplet defect systems over the number of unique defect systems, i.e., 403/627 ∼ 64%. These two ratios demonstrate that the symmetry-enabled strategy based on site symmetries dramatically enhances the possibility of finding defects with triplet ground states. In addition, the results show that 58% of triplet defects are antisites and the rest (42%) are vacancies. The fractions of charge states of 1-, 0, and 1 for antisites and vacancies are 26%, 16%, and 16% and 13%, 19%, and 10%, respectively (see Supplementary Table 3).

Intra-defect transitions are crucial to the operation of a quantum defect. As the next screening layer, we select those triplet defects by the criterion that the energy gap between the highest occupied defect level (HODL) and the lowest unoccupied defect level (LUDL) must be greater than or equal to 0.5 eV in at least one spin channel. To account for the potential underestimation of band gaps from the r$^2$SCAN functional, we choose to keep those defect systems with defect levels that are submerged into the band edges by less than 0.25 eV. 152 triplet defects satisfy the criterion and 90% of the defects following the criterion are antisites, indicating that antisite is a dominant type of quantum defects in binary 2D materials. The distributions of charge states -1, 0, and 1 for antistes and vacancies are 39%, 31%, and 20% and 5%, 2%, and 4%, respectively (see Supplementary Table 4). Note that the formation of vacancy complexes may create defect level splitting by breaking local symmetries and therefore increase the energy gaps between occupied and unoccupied defect levels. This defect complex design strategy deserves a separate study but is out of the scope of our current work.

The triplet defects from Supplementary Table 4 are created based on another round of high-throughput computations using the hybrid functional (HSE06) to mitigate the potential band gap issue for semiconductors. Taking into account of uncertainty in the computational results, we choose to keep those defect systems with defect levels that are submerged into the band edges by less than 0.5 eV. Based on the HSE06 results, 141 triplet spin defects are identified. Among them, there are 113 host materials accounting for nearly 50% of the initial material set (235 host materials). This again demonstrates the power of site-symmetry-based guidance when searching for host material candidates and suitable crystal sites for creating potential triplet spin defects (see Supplementary Table 5).

For the triplet defects with energy gaps between HODL and LUDL larger than or equal to 0.5 eV, we perform a set of calculations including constrained DFT (CDFT) for the energy of ZPL and the corresponding Franck-Condon relaxation energies. The transition dipole moments from wavefunctions are calculated to estimate the dipole transition rates. In addition, the ZFS of sublevels in the ground state due to spin-spin dipolar interaction is also calculated (see Methods for details). With these parameters related to quantum defects, we are able to find promising quantum defect candidates based on the criteria related to magneto-optical properties. The criteria for optical transitions are chosen as: (i) Wavelength of the ZPL must be shorter than or equal to 2500 nm (0.496 eV), corresponding to telecom wavelength. (ii) Parameter $D$ of the ZFS related to the robustness and operation of qubits is larger than or equal to 0.95 GHz (L-band) (iii) Intra-defect transitions must be dipole-allowed. (iv) The energy difference between HODL and VBM and that between CBM and LUMO must be at least 0.095 eV.

There are 45 charged defects following the criteria and forming a set of quantum defects (see Supplementary Table 7) Those defects are comprised of 41 host materials, including 12 post-transition-metal monochalcogenides (PTMCs), four transition-metal dichalcogenides (TMDs), 20 halogenides, and other compounds. The summarized workflow is shown in Figure 1b, and the statistics of these qubit-related parameters will be discussed in the following section.

**Optical transitions and sublevel splittings**

The properties of optical transitions, including ZPLs and the corresponding Franck-Condon relaxation energies for the quantum defect candidates, are crucial to the applications of emitter and qubit operations. The distribution of ZPL wavelengths is shown in Figure 2a. Four candidates (9%) show intra-defect optical transitions in the range of visible light (380-780 nm), and 43 candidates (96%) may be involved in optical transitions in the range of near-infrared light (NIR) (700-2500 nm). Among those optical transitions within the NIR range, we further categorize them by the range of biological imaging (650-950 and 1000-1350 nm)[47] and telecommunication operating wavelength for optical fibers (1260-1565 nm). There are 28 candidates (62%) that may create optical transitions in the spectrum range of biological imaging and 11 candidates (24%) that may generate transitions in the spectrum range of telecommunication.

In addition to the ZPLs, the Franck-Condon relaxation energies of both ground and excited states are also important for understanding phonon side-bands in the spectrum. The relaxation energy distribution of candidates is shown in Figure 2b, where the notations BC and DA represent the relaxation energies of the triplet excited state and the triplet ground state, respectively. The other notations AB and CD shown in the right-hand-side diagram of Figure 2b represent the vertical transitions from a ground state to an excited state and *vice versa*. The relaxation energies of the excited states in 39 defect candidates (87%) are less than or equal to 0.05 eV and those of the ground states in 42 candidates (93%) are less than or equal to 0.05 eV. These results unveil a general observation that the ZPL emission for most antisite defect candidates is expected to be strong and sharp.

The quantum efficiency, defined as the ratio of radiative transition rate and total transition rate (sum of radiative and nonradiative rates), is another crucial factor for evaluating the performance of quantum emitters. In this work, we symmetrically evaluate the radiative transition rates by calculating the transition dipole moments (see Methods for details). The results show that 14 candidates (31%) have radiative rates larger than or equal to 6 MHz, the measured radiative transition rate from a set of emitters in h-BN.[48-50] Note that those emitters in h-BN have a relatively low quantum efficiency of 6% ~ 12%.[51,52] The proposed defect systems with radiative rates larger than 5.98 MHz could potentially provide greater quantum efficiency (Figure 2c). Noted that singlet states in these localized defects are usually highly correlated. The theoretical treatment of these singlet states and the computations of nonradiative rates in a high-throughput manner is beyond the scope of current work.

The realization of a defect qubit is based on implementing multiplets of the ground state as qubit states. Since these qubit candidates have triplet ground states, the spin Hamiltonian consists of the Zeeman term and the spin-spin dipolar interaction. Without applying an extra magnetic field, the predominant interaction that causes the splitting of the sublevels in the triplet ground state is the spin-spin dipolar interaction. Note that the spin-orbit coupling can also cause the splitting of sublevels, while the small effective orbital angular momentum in a ground state suppresses this effect.[53] For this reason, we assume that the spin-spin dipolar interaction is dominant and calculate this interaction to evaluate the splitting of multiplets in the ground state. To calculate the spin-spin dipolar interaction, we calculate the parameter $D$ of ZFS, which evaluates the energy gap between $m_s = \pm 1$ and $m_s = 0$ in a triplet ground state (see Methods for details). The ZFS $D$ is related to qubit coherence time, especially the spin-lattice relaxation time $T_1$. The higher $D$ is, the more robust the qubit is within the microwave operation range. The results show that $D$s of the candidates are in the frequency range from 1 GHz to 12 GHz, corresponding to L-, S- C-, and X-band according to standard electrical engineering definitions (see Figure 2d).

**Discussion**

**The relationship between ZFS and defect state localization**

The ZFS *D* is a measure of the splitting of energy levels due to spin-spin dipolar interaction. This interaction can be calculated using the formula: $D_{ab} = \frac{1}{2}\frac{\mu_0}{4\pi}\frac{g_e^2\mu_B^2}{S(2S-1)}\sum_{i>j}^{occupied}\eta_{ij}\int |\Phi_{ij}(r_1,r_2)|^2 \frac{r^2\delta_{ab}-3r_ar_b}{r^5}dr_1dr_2$, where $\Phi_{ij}$ is the Slater determinant of Kohn-Sham orbitals *i* and *j*, and $\eta_{ij}$ is 1 or -1 for KS orbitals within the same or different spin channels (see Methods). This formula suggests that in general localized orbitals experience strong interactions due to their inverse relationship with distance.

Motivated by this formula, we examine the relationship between the ZFS *D* and the atomic radius of antisite species and find that there is a clear inverse relationship, as shown in Figure 3a. The atomic radius of an antisite species can be viewed as the extent of localization of defect states, as these states are largely contributed by the antisite. As the atomic radius of the antisite species increases, the localization of defect levels decreases, leading to a weakening of spin-spin dipolar interaction.

To further evaluate the spatial localization of these states, we calculate the inverse participation ratio (IPR) of the highest occupied defect levels in the spin-up channel for the antisite defects. Since the antisite defects have a triplet spin state, the defect-level configurations have two types: 1-2 and 2-1 splitting. Additionally, since the number of spin-up electrons is greater than the number of spin-down electrons in the candidates, and DFT treats spin channels independently, we choose the highest occupied defect levels in the spin-up channel to calculate the IPR. The result shows that the ZFS D generally increases as the spin-up HODL IPR increases, supporting our earlier conclusion (see Figure 3b and 3c).

**Estimation of ZPL with crystal field theory**

ZPL is an important parameter in the study of quantum defects, particularly in relation to quantum emitters and qubit operations. The results of optical transitions indicate that ZPL is primarily determined by the vertical transitions and Franck-Condon relaxation energies in the configuration diagram of the defects. The vertical transition is correlated with the size of the splitting between LUDL and HODL, making it possible to qualitatively estimate ZPL by considering the splitting between LUDL and HODL. Using the crystal field theory (CFT) commonly applied in coordination compounds, we estimate the splitting of LUDL and HODL by taking into account the Coulombic interactions between an antisite and its adjacent atoms. CFT posits that the splitting of $d$-orbitals in transition metals in coordination compounds can be evaluated by the Coulombic interactions between the metal ion and its adjacent ligands, assuming that all ions are point charges in the compounds. It is important to note that CFT is a simplified model that does not account for other bonding interactions between the metal ion and ligands. Despite this simplification, CFT can still be used as a rough evaluation of the interactions between a metal ion and ligands.

The technique of Bader charge analysis is used to verify the charge of an antisite and its adjacent atoms, which have the same chemical species.[54] We then examine the relationship between the ZPL and Coulombic interactions in 45 candidates. The Coulombic interactions are calculated by using the potential energy equation $\frac{q_i q_{antisite}}{r_i}$, where $q_i$, $q_{antisite}$, and $r$ are the Bader charges of the adjacent atom $i$, the antisite, and the distance between the adjacent atom $i$ and the antisite, respectively. We take the average value of the potential energies among all pairs of an antisite and its adjacent atoms in the candidates and then take the absolute value of this average to focus solely on the strength of the Coulombic interaction (see Supplementary Figure 2a).

To understand the relationship between the zero-phonon line (ZPL) and Coulombic interactions, we categorize the candidates by the chemical group of the antisite species. Our analysis shows that four groups (VIB, IIIA, IVA, and VIIA) have a positive relationship between ZPL and Coulombic interactions, which indicates that the crystal field theory (CFT) is able to explain the splitting of LUDL and HODL in these groups (see Supplementary Figure 2b, 2d, 2e, and 2g). However, the chemical groups IIB and VIA shows a negative relationship between ZPL and Coulombic interactions (see Supplementary Figure 2c and 2f). This may be due to the fact that CFT is not sufficient to explain the bonding interactions in these antisite defect systems and more complex bonding interactions need to be considered.

**The relationship between wavefunction localization and spin state energy differences**

As mentioned in the symmetry-enabled strategy, the exchange interaction between electrons with the same spin stabilizes spin-triplet defects. This interaction between a pair of electrons in orbitals $\psi_i$ and $\psi_j$ can be defined as $\int \psi_i(r_1)^\star \psi_j(r_2)^\star \frac{1}{|r_1-r_2|} \psi_j(r_1)\psi_i(r_2) dr_1 dr_2$, where $r_1$ and $r_2$ are the positions of electron 1 and 2. Localized orbitals experience strong exchange interactions because the denominator of this formula is inversely related to the distance between the electrons. In addition, the on-site interaction between the spin-up and spin-down electrons at an orbital is enhanced for spin-singlet defects when the occupied orbital becomes more localized. With these mechanisms, the spin state singlet becomes less favorable compared to the spin state triplet of a defect (see Supplementary Figure 3g).

We then examine the relationship between the total energy difference of a defect's spin state (singlet and triplet) and the summed spatial localization of the defect's occupied levels in both spin states (see Supplementary Figure 3a). The spatial localization of a defect wavefunction is evaluated using the IPR. The results show that as the total localization increases, the energy difference between spin states also increases due to the enhancement of exchange and on-site interactions. We also group the candidates by the chemical groups of the antisite species. A positive relationship between the total localization and energy difference between spin states is observed for most groups (see Supplementary Figure 3b-7f).

This observation indicates that the energy difference between spin states is enhanced by the stronger exchange and on-site interactions as the levels become more localized. In this study, we only consider the relationship between the exchange and on-site interactions and the total energy difference between spin states. It should be noted that more complicated correlation interactions may also be involved in stabilizing spin-triplet defects.

**Qubit operation principles of defect qubit candidates**

To illustrate a complete loop of qubit operations, identifying singlet states and intersystem crossings between triplets and singlets are essential. Singlet spin states are highly correlated, which can be calculated by advanced computational methods such as many-body perturbation theory (MBPT). Meanwhile, the electronic configuration based on symmetry renders a qualitative way to analyze singlet states and intersystem crossings. The numbers of defect candidates consisting of symmetries $C_{3v}$ and $D_{2d}$ are 34 and 11, respectively. Based on symmetry analysis, we address the electronic configurations of these defect qubit candidates.

In the discussions above, two types of defect-level splittings (1-2 and 2-1) are formed for these defect candidates. To simplify symmetry analysis of term symbols, sublevels, and intersystem crossings, we adopt the duality relationship between electrons and holes to replace the 1-2 type electronic configuration with the 2-1 type hole configuration. That unifies the configurations as 2-1 type half-occupied doubly degenerate levels $e^2$. Note that the IRs of unoccupied single levels are not identical for $C_{3v}$ and $D_{2d}$. We deduce the term symbols for triplet ground states and singlet states of defect candidates by decomposing the tensor-product representation of $e \otimes e$ and following Pauli's exclusion principle. The results show that $C_{3v}$ and $D_{2d}$ have the identical term symbol $^3A_2$ for the triplet ground state. For singlet states, $C_{3v}$ has $\{^1A_1, ^1E_1, ^1E_2\}$ and $D_{2d}$ has $\{^1A_1, ^1B_1, ^1B_2\}$. $E$ is a 2D IR, and thus $^1E_1$ and $^1E_2$ are degenerate. It is known that ordering singlet states is critical for qubit operations. Hund's rule states that, due to the antisymmetric spinor of singlet states, the state energy is lower when the angular momentum of the state is larger. Based on this rule, $^1E$ is more stable than $^1A_1$.[55] The singlet states from $D_{2d}$ are formed by $\{^1A_1, ^1B_1, ^1B_2\}$, where $^1B_1$ and $^1B_2$ have identical wavefunctions as $^1E_1$ and $^1E_2$ from $C_{3v}$ (see Supplementary Table 9), $^1B_1$ and $^1B_2$ belong to two different 1-dimensional (1D) IRs. Therefore, they are not essentially degenerate. Although the positions of singlet states cannot be acquired directly by DFT, we can approximate it by considering the matrix element of Coulombic interaction. The energy separation between $^3A_2$ and $^1E_1$ are equal to those between $^1E_2$ and $^1A_1$ for defects with $C_{3v}$. Because $^1E_1$ and $^1E_2$ are degenerate, we can approximately estimate the ordering of three singlet states as $\{^3A_2, ^1E_1/^1E_2, ^1A_1\}$. Followed by the same argument, we show that the energy separation between $^1A_1$ and $^1B_1$ is equal to that between $^1B_2$ and $^3A_2$ for defects with $D_{2d}$. Since $^1B_1$ and $^1B_2$ are not necessarily degenerate, the ordering of $^1B_1$ and $^1B_2$ must be determined by explicit calculations. Thus, the approximate ordering of singlet states is {$^3A_2$, $^1B_2$, $^1B_1$, $^1A_1$}, based on the assumption that $^1B_2$ is more stable than $^1B_1$.

Sublevels of triplet states and singlet states are also crucial to qubit operations. Here we label all sublevels with IRs and identify symmetry-allowed intersystem crossings. The splitting between sublevels $m_s = 0$ and $m_s = \pm1$ in triplet ground states is also evaluated by ZFS $D$ as discussed in the previous section and shown in Supplementary Table 7. The sublevel formation in triplet excited states is beyond our scope. The related quantities in NV center has been computationally studied.[56]

Intersystem crossings are essential for initialization and readout processes as part of qubit operations. Assuming that the spin-orbit interaction $H_{so}$ largely mediates intersystem crossings, whether an intersystem crossing is allowed can be determined by symmetry analysis. Since $C_{3v}$ and $D_{2d}$ are axial symmetries, $H_{so}$ can be defined as $H_{so} = \sum_k \lambda_{xy}(l_k^+ s_k^- + l_k^- s_k^+) + \lambda_z l_k^z s_k^z$, where $l_k^\pm$ and $S_k^\pm$ are the raising and lowering operators for the angular momentum and spin operator, respectively.[57] $H_{so}$ can be decomposed into an axial part $H_{so}^\perp$ and a nonaxial part $H_{so}^{//}$. $H_{so}^\perp$ changes the electronic or hole configurations and the spin projected quantum numbers due to the angular momentum and spin raising and lowering operators. On the other hand, $H_{so}^{//}$ preserves those quantities as it is only composed of the $z$-component angular momentum and spin operators. Accordingly, we identify the allowed intersystem crossings for each symmetry in Figure 4a, 4b, and Table 1.

The complete loop of qubit operations consists of three steps, initialization, manipulation, and readout. Initialization is to polarize a qubit into a "zero" qubit-state such as the sublevel $m_s = 0$ in the triplet ground state. Then, by optically pumping a qubit system into its triplet excited state, the intersystem crossings between triplet and singlet states allow a qubit in an initial state to relax back to the "zero" qubit-state. Manipulation is to utilize zero- and one-qubit states defined as the sublevels $A_1$ ($m_s = 0$) and $E$ ($m_s = \pm1$) of the triplet ground state, which is implemented by applying resonant microwave. Finally, readout is to retrieve the outputs after manipulation, which can be achieved by detecting optical fingerprints such as the intensity difference of the optical transitions from the sublevels in excited states to zero- and one-qubit states in ground states. The intensity difference is due to the portion of the population that can relax back to the ground state via intersystem crossings, resulting in a dimmer intensity of the fluorescence.

The schematic diagram of the complete loop of qubit operations is shown in Figure 4c. Notice that we identify that two vacancies $V_{Al}^-$ and $V_B^-$ with $D_{3h}$ symmetry from the graphene-like AlN and h-BN host triplet ground states. Negatively-charged aluminum vacancy $V_{Al}^-$ in AlN does not provide any allowed intersystem crossing from triplet excited state to singlet state. It is thus excluded from the candidate set. On the other hand, negatively-charged boron vacancy $V_B^-$ in h-BN is a well-known and promising defect qubit candidate.[32,58] Previous computational work has studied this qubit system in detail on the higher theory level and revealed a more complex energy diagram.[32,58]

**Antisites in post-transition-metal monochalcogenides: an outstanding quantum defect family**

Among the defect qubit candidates we proposed, antisite defects in PTMCs stand out from other candidates. As a novel class of 2D materials, PTMCs are composed of metals from group III to V (M: Ga, In, Ge, Sn, Sb, and Bi) and chalcogens (X: S, Se, and Te) with structures of 4 stacked sublayers in the order of X-M-M-X bonded by van der Waals interactions. Group-III PTMCs have attracted attention in recent years due to their supreme electronic properties, including thickness-dependent band gaps and high electron mobility.[41-45] Besides, monolayers of PTMCs have been synthesized by mechanical exfoliation.[45,59-62] Both intrinsic and extrinsic point defects have been observed in GaSe and InSe, including cation vacancies, anion vacancies, and oxygen impurities on anion sites.[63-65] The quantum defects predicted in this work are based on group-III MX of PTMCs where metals M are Ga and In.

Here we focus on a set of antisite defects in group-III MX with two space groups *P-6m2* and *P-3m1*, which can resemble H- and T-phase in TMDs. Anion and cation sites in both phases have the same local symmetry $C_{3v}$, because the local motifs of upper two X-M sublayers and lower two M-X sublayers are identical. Anion antisite $M_X^{1+}$ (X: S, Se, M: Ga) forms a triple ground state with a 2-1 type defect-level splitting. Cation antisite $X_M^{1-}$ (M: Ga, In; X: S, Se) also forms a triplet ground state with a 1-2 type splitting (see Supplementary Figure 6). The structural parameters for both pristine and defective structures are shown in In Supplementary Table 10 and 11. Additionally, Supplementary Table 12 shows their electronic structure related properties.

There are several advantages of utilizing these antisites as qubits: (i) a set of well-isolated in-gap levels are available for qubit operations; (ii) a wide range of ZPLs and ZFS $D$s can be achieved by selecting host materials; (iii) mature host materials are available experimentally for defect implementation. The minimum energy gaps between the HODL and the VBM and between LUDL and CBM of the antisites in group-III PTMCs are 0.218 eV and 0.098 eV, respectively, showing the defect levels are well-preserved in the band gaps. The weakest ZPL among the antisites is 0.882 eV, which are operable in the wavelength region of telecommunication, and the ZFS $D$s are in the operational range of microwave (see Supplementary Table 13). Since these antisite defect systems are composed of $s$- and $p$-elements, the decoherence effect from spin-orbit coupling can be suppressed, compared to anion antisite qubits in TMDs dominated by heavy transition metal elements. This large set of antisite defects in group-III PTMCs shows promising potential for QIS applications such as qubits and emitters.

Since the triplet states are formed in the anion antisites with a 1+ charge state and the cation antisites with a 1- charge state, the stability of these charge states is an important factor for realistic applications. Therefore, we calculate the transition levels of these charged antisite defects. The results show that the Fermi-level windows of the 1- charge states in the cation antisites (shown in Supplementary Figure 7 by the red double-arrow lines) range from 0.88 eV to 1.44 eV. In contrast, those of the 1+ charge states in the anion antisites (shown in Supplementary Figure 7 by the purple double-arrow lines) have a smaller range from 0.31 eV to 0.45 eV (see the column "$E_F$ window" in Table 8). Note that an interpretation of transition levels is that the Fermi level positions at which the formation energies of the defects in two distinct charge states are equal. And a given defect system can be stable in various charge states depending on the positions of the Fermi level (see Methods for details).

Note that the substitution of these antisite defects by other elements in the same group may host similar properties, hence greatly expanding the candidate set.

**Summary**

Point defects hosting triplet ground states are desirable for quantum information science and technologies. We demonstrate that the symmetry-enabled strategy combined with data-driven techniques can significantly enhance the probability of finding triplet defects. Multiple important magneto-optical properties, including dipole transition rate, ZPL, and ZFS are calculated and used for the data-driven discovery of quantum defects. We show that 45 antisite defects satisfy the material criteria for quantum information technologies such as qubits and quantum emitters.

This comprehensive dataset of point defects in 2D materials provides information about the candidates that allows us to investigate the relationships between observables such as ZFS and ZPL, and parameters such as atomic radius, spatial localization of levels, and crystal-field splitting. We also demonstrate that the energy difference between spin states is correlated with the exchange and on-site interactions, indicating that the interactions play a crucial role in stabilizing a spin-triplet defect. This finding is also consistent with the symmetry-enabled strategy.

The operational principles of qubit candidates are discussed based on the identified optical transitions and intersystem-crossing paths. We propose 16 anion antisites and cation antisites in group-III PTMCs as a novel quantum defect design platform. With the well-defined in-gap defect levels, optimal magneto-optical properties, and the availability of PTMCs as well-studied 2D semiconductors, this set of 16 antisite defects has a great potential to becoming promising quantum defects for technical applications, including qubits, quantum emitters, and sensors.

## Methods

**Full workflow of implementing the symmetry-enabled strategy**

To summarize the defect screening workflow (Figure 1b), the host materials from the C2DB database are selected based on the following requirements: (1) nonzero band gap calculated by hybrid functional (HSE06) without spin-orbit coupling, (2) nonmagnetic ground states, and (3) binary compounds. Subsequently, the selected hosts are calculated using the SCAN functional for their basic material properties, including the optimization of crystal structures, the density of states, and band structures with band IR information. The hosts go through another screening layer based on the criteria: (1) band gap larger than or equal to 1 eV; (2) the crystal structure satisfies the symmetry-enabled strategy showing 2D or >2D IRs. The defect structures are created and calculated with the r$^2$SCAN functional for the optimized structures and electronic structures. The defects with triplet ground states are further screened so that the energy gaps between LUDL and HODL are greater than or equal to 0.5 eV. Based on the retrieved information of triplet spin defects from the above process, we perform the defect calculations based on HSE06 functional to accurately describe defect levels and bulk band gaps. Notice that the defect structures are built based on the host structures in the C2DB database. The resulting triplet spin defects are further screened by using the criterion that the energy gap between LUMO and HODL is greater than or equal to 0.5 eV. Fundamental material properties related to these quantum defect candidates, including ZFS, ZPL, and dipole transition rates, are calculated to identify their potential for quantum information technology applications, including qubits and quantum emitters. The criteria include: (i) The energy differences between HODL and VBM and between CBM and LUMO must be at least 0.095 eV. (ii) The wavelength of ZPL must be shorter than or equal to 2500 nm (0.496 eV). (iii) The intra-defect transitions must be dipole-allowed. (iv) The ZFS $D$ is larger than or equal to 0.95 GHz.

**Computational details**

All calculations were performed by using the Vienna Ab initio Simulation Package (VASP)[66] based on the density functional theory (DFT)[67,68]. To calculate the spin density near the nuclei, the projector-augmented-wave method (PAW)[69,70], and a plane-wave basis set were used. In the workflow, we adopted SCAN[71] and r$^2$SCAN[72] metaGGA functionals for hosts and defects, and then the screened hybrid-functional of Heyd-Scuseria-Ernzerhof (HSE)[73,74] with default mixing parameter and the standard range-separation parameter (0.2 Å$^{-1}$) were used for overcoming the well-known band-gap problem of the traditional DFT. For defect supercell calculations, we used the **Γ** point in the Brillouin zone for defect-state calculations to avoid undesirable splitting of defect states. A vacuum space of 20 Å was added along the direction perpendicular to the monolayer with an appropriate size of planar supercell with the distance between image defects at least 15Å, in order to avoid interactions between the adjacent images. Structural relaxations have been performed for all the systems investigated which were converged until the force acting on each ion was less than 0.01 eV/Å. The convergence criteria for total energies for structural relaxations and self-consistent calculations are 10$^{-4}$ eV and 10$^{-5}$ eV, espectively.

The constrained DFT (CDFT) methodology[75,76] was employed for the calculation of excitation energies between the triplet states. Dipole transition rate is calculated from Fermi's golden rule[77,78] $\Gamma_{rad} = \frac{nE_{ZPL}^3\mu^2}{3\pi\epsilon_0 c^3\hbar^4}$, where $E_{ZPL}$ is ZPL energy, $\mu$ is transition dipole moment. The refractive index here is set as 1 for 2D materials. ZFSs of the sublevels in the triplet ground states due to the spin-spin dipolar coupling were evaluated by spin-spin dipolar Hamiltonian, $\hat{H}_{SS} = \hat{\mathbf{S}}\mathbf{D}\hat{\mathbf{S}}$, where $\hat{\mathbf{S}}$ is the total-spin operator and **D** is a 3×3 ZFS tensor.[79] For symmetries $C_{3v}$ and $D_{2d}$, $\hat{H}_{SS}$ can be expressed as $\hat{H}_{SS} = D\left(S_z^2 - \frac{S(S+1)}{3}\right) + E(S_x^2 - S_y^2)$ where $D = \frac{3}{2}\mathbf{D}_{zz}$ can be measured in experiment. $S_{x,y,z}$ is the spin projected on $x$-, $y$-, and $z$-direction, and $S$ is the total spin. The scalar parameters $E$ defined as ($\mathbf{D}_{xx}$-$\mathbf{D}_{yy}$)/2 describes the splitting between the sublevels $m_s$= 1 and $m_s$ = -1. The modified Python package Pyzfs[80] was used to calculate the scalar ZFS $D$ within the framework of PAW-ps used by VASP, which has yielded reasonable results for the NV$^-$ center.[81,82]

**Defect formation and transition levels.**

Relative stability of point defects depends on the charge states of the defects. We analyzed the stability of antisite defects in PTMCs by calculating the defect formation energy ($E_f$) for charge state q, which is defined as: $E_f(\epsilon_F) = E_{tot}^q - E_{bulk} + \mu_X - \mu_M + q(\epsilon_F + E_V) + \Delta E$, where $E_{tot}^q$ is the total energy of the charged-defect system with charge $q$, $E_{bulk}$ is the total energy of the perfect MX system, $\mu_M$ is the chemical potential of the metal atom M, $\mu_X$ is the chemical potential of the anion atom X, $\epsilon_F$ is the position of the Fermi-level with respect to the valence band maximum $E_V$, and $DE$ is the charge-correction energy. Transition levels are defined as $\epsilon(q'/q) = (E_f^{q'}-E_f^{q})/(q-q')$, where $E_f^q$ is the formation energy for the state of charge $q$. We can interpret the transition levels or ionization energies as the Fermi-level positions at which the lowest-energy charge state of defect changes. In a low-dimensional system, due to anisotropic screening, ionization energy (IE) diverges with respect to the vacuum, and we applied ,a charge-correction method.[83,84] We assume that the chemical potential of M and X are in thermal equilibrium with $MX_2$, i.e., $\mu_{MX} = \mu_M + \mu_X$, where $\mu_{MX}$ is the energy of the perfect MX system. The accessible range of $\mu_M$ and $\mu_X$ can be further limited by the lowest energy phases of these elements depending on growth conditions. It should be noted that the transition levels do not depend on the choice of chemical potentials.

**Cross-validation of defining defect levels by IPR and projected DOS.**

Identifying defect levels from a calculated defect system can be difficult, especially for those defect levels that are close to band edges. A direct approach to identifying defect levels is by visualizing the spatial distribution of the wavefunction of defect levels. It is identified as a defect level if the distribution of the wavefunction is localized around the defect. Although this direct approach is straightforward, the visualization of wavefunctions is time-consuming and thus not suitable for high-throughput calculations. As a result, finding parameters that can be used to evaluate the extent of the localization of an orbital is required. As a defect level is localized around defects, we can extract the site- and orbital-projected wavefunction characters of each orbital and then sum the characters over the nearest neighbors of a defect. Notice that VASP software generates an output file PROCAR, storing such information. The threshold of the summed site-projected character is set to 0.2 to determine defect levels for the majority of defects in the results of high-throughput calculations. An alternative approach to defining localized defect levels is to adopt the inverse participation ratio (IPR). It is defined as $IPR(\psi) = \sum_{n=1}^{N}|\psi(n)|^4$, where $\psi$ is a normalized wavefunction and $n$ iterates over $N$ grid points in real space. The IPR of 1 for an orbital indicates the orbital is completely localized at a grid point in real space. We show that both approaches can be used to determine defect levels and give identical results by taking the defects $Te_{Al}^{1-}$ in $Al_2Te_2$ and $S_{Ga}^{1-}$ in $Ga_2S_2$ as examples (see Supplementary Figure 1).

**Acknowledgement**

This work is supported by the National Science Foundation under Grant No. 2314050. This research used resources of the National Energy Research Scientific Computing Center (NERSC), a U.S. Department of Energy Office of Science User Facility located at Lawrence Berkeley National Laboratory, operated under Contract No. DE-AC02-05CH11231 using NERSC award BES-ERCAP0029544.

**Author contributions statement**

J.Y.T. and Q.Y. conceived the experiments. Z.F. wrote the workflow for high-throughput computations and quantum defect screening, J.Y.T. and W. G. performed data analysis. Q.Y. supervised the study. J.Y.T. wrote the manuscript and all authors contributed to the revision of the manuscript.

**Competing interests**

The authors declare no competing interests.

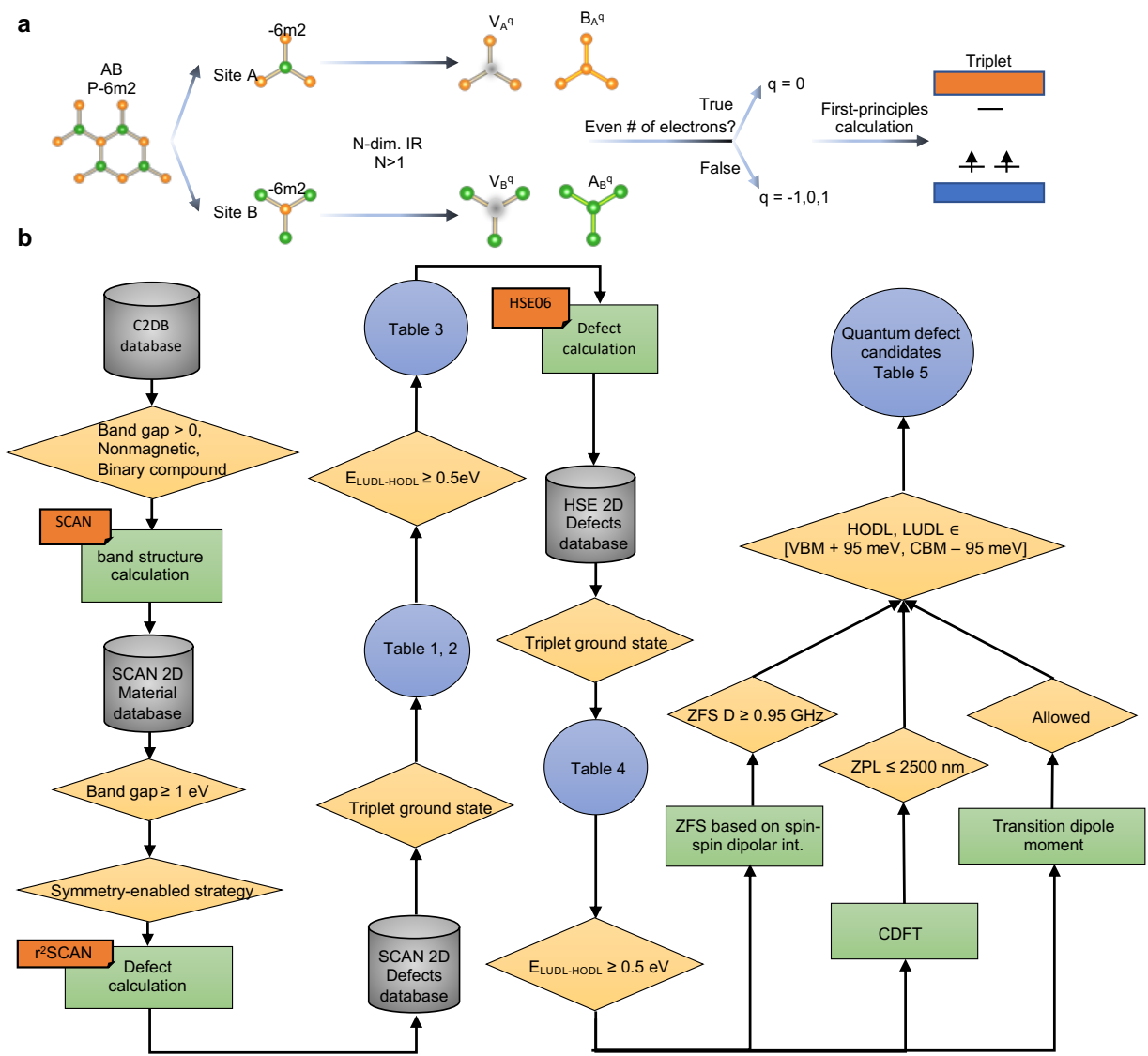

Figure 1 **Workflows of creating defects and performing calculations based on the symmetry-based strategy. a** A schematic workflow of creating defect structures based on the site-symmetry analysis. The charge states are determined by the number of electrons. **b** A full workflow of symmetry-enabled discovery of quantum defects. The metallic color cylinders represent the databases for storing inputs and outputs of the calculations. The green and the smaller orange rectangles represent the performed calculations and the used functionals. The purple-blue circles denote the tables shown in the main text. Finally, the yellow parallelograms represent the applied criteria for the screening.

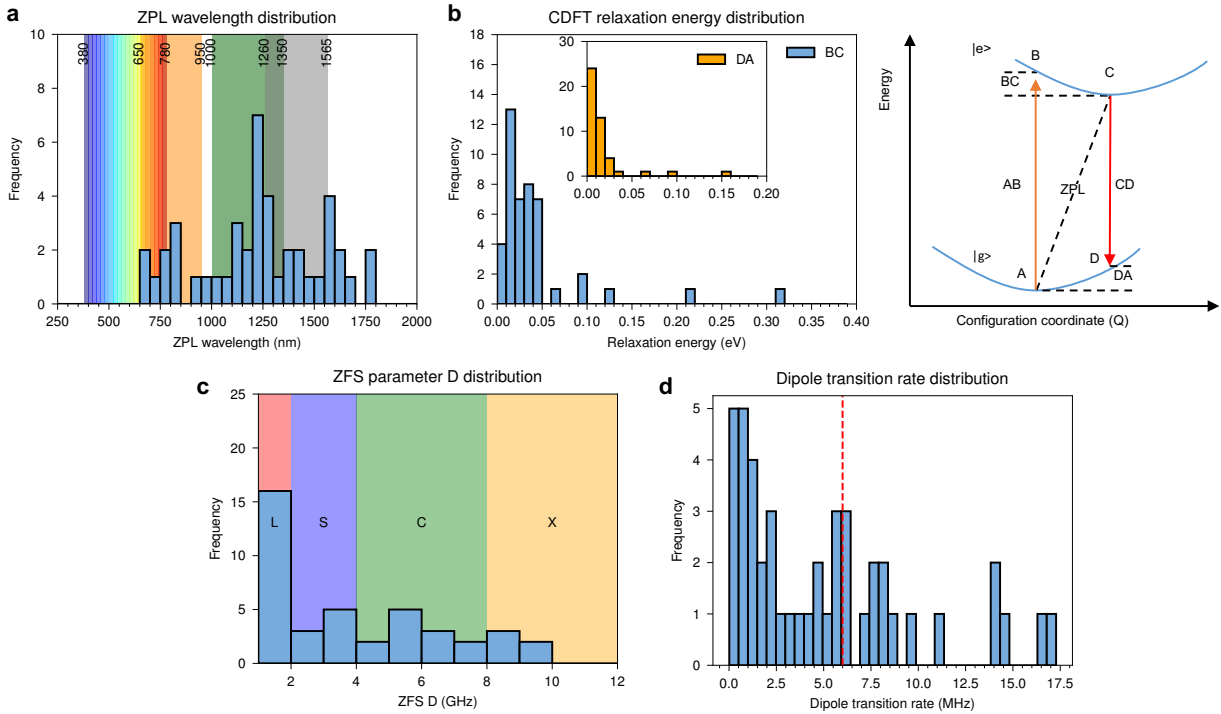

Figure 2 **Statistics of qubit-related properties for candidates from , including intra-defect optical transitions and ZFS D in their ground states. a** A histogram of ZPL wavelength for quantum defect candidates. The rainbow-color block represents the visible-light spectrum. The orange and green blocks represent the wavelength range for biological imaging. The grey block represents the telecommunication operation wavelength for optical fibers. **b** A histogram of relaxation energies of triplet excited state (in blue) and triplet ground state (in orange in the inset). **c** A histogram of dipole transition rates. Two groups of transition-rate distributions are segmented by 6 MHz, the measured radiative transition rate from a set of emitters in h-BN (red dashed line). **d** A histogram of the ZFS $D$ due to spin-spin dipolar interactions. The pink, violet, green, and yellow blocks represent L-, S-, C-, and X-band.

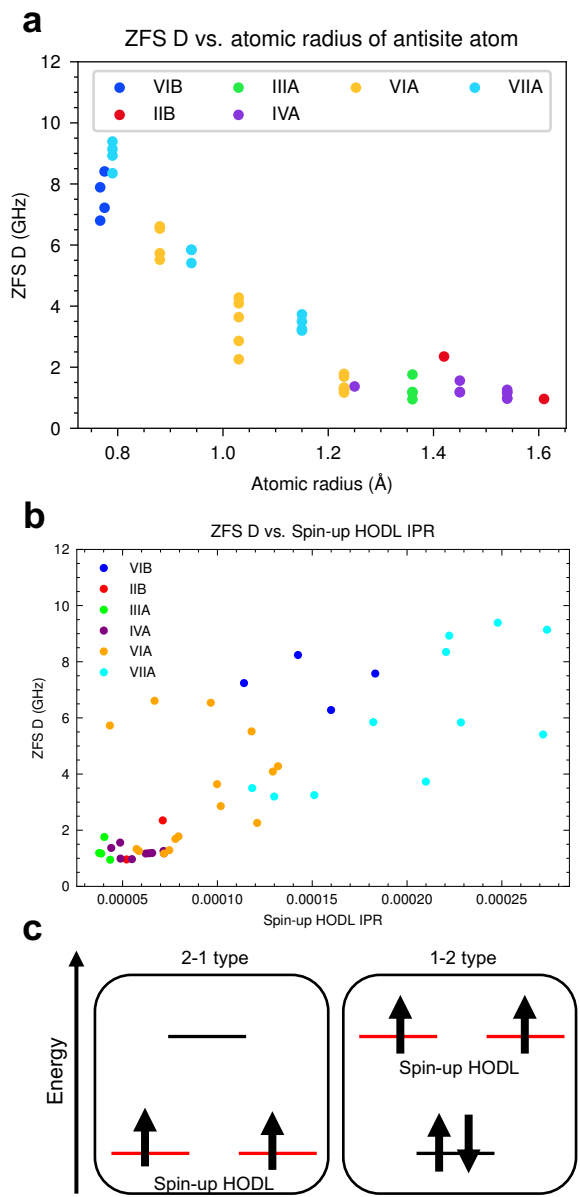

Figure 3 **Relationships between ZFS *D* and atomic radius and ZFS *D* and spin-up HODL IPR for 45 quantum defect candidates.** **a** The inverse relationship between ZFS *D* and atomic radius of the antisite species. The colors denote the group of the periodic table for the antisite species. Notice that the cationic radii are chosen for the group-VIB antisites. **b** The relationship between ZFS *D* and spin-up HODL IPR for the antisite species. The spin-up HODL IPR is defined as the IPR of the highest occupied defect levels in spin-up channel in the 2-1/1-2 type of LUDL-HODL splitting types (red lines in the **c**), and the schematic illustration is shown in **c**.

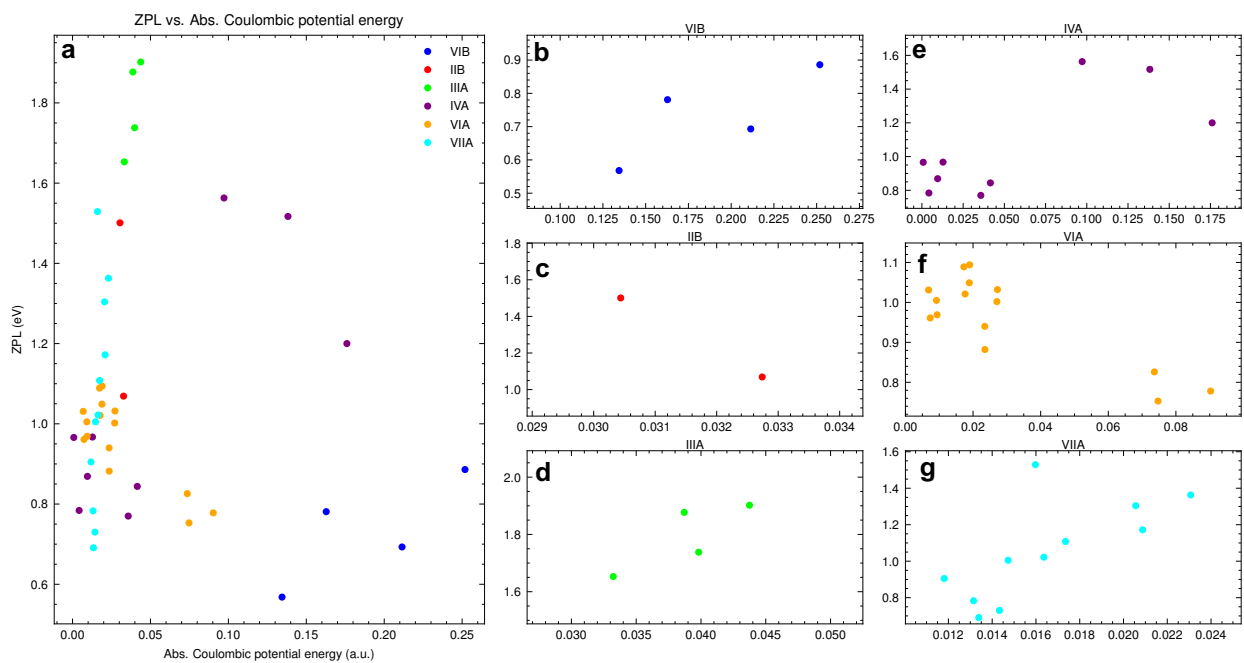

Figure 4 **Estimation of ZPL based on crystal-field splitting between LUDL and HODL for 45 quantum defect candidates.** The splitting of LUDL and HODL affects a ZPL of defects, and we can quantitatively estimate the splitting by the Coulombic interaction between an antisite and its adjacent atoms. The Coulombic interactions between an antisite and its adjacent atoms are averaged out, and we only focus on its strength of it. **a** The correlation between ZPLs and interactions for 45 candidates. The color dots represent various chemical groups of the antisite species. **b-g** The sub-plots of the plot **a** that are categorized by the chemical groups are presented. As shown, most of the chemical groups show a positive correlation, while the sub-plots **c** and **f** show a negative correlation.

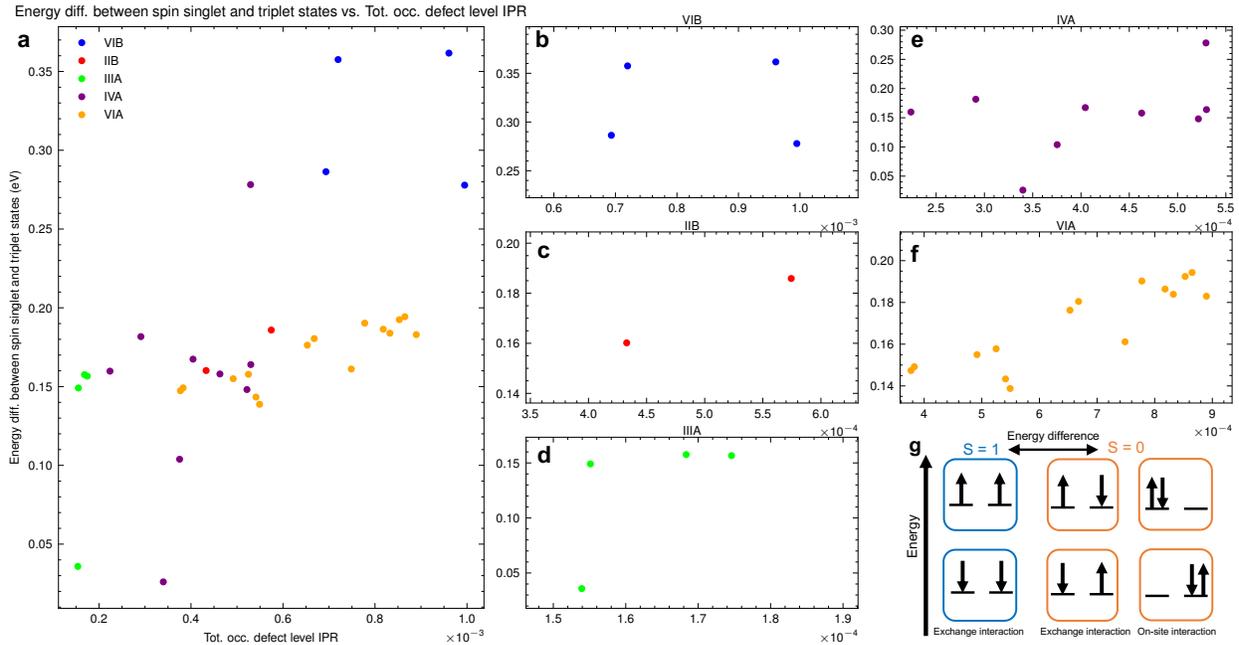

Figure 5 **Correlation between total energy difference of a defect's spin state (singlet and triplet) and the summed spatial localization of the defect's occupied levels in both spin states.** We show that the exchange and on-site interactions contributing to the total energy difference between spin-singlet and -triplet defects are correlated with the summed localization of occupied defect levels in both spin states. **a** The correlation between the energy difference of spin-singlet and -triplet and the summed localization of occupied defect levels in both spin states. The color dots represent various chemical groups of the antisite species. **b-f** The sub-plots of **a** that are categorized by chemical groups are presented. **g** A schematic diagram illustrates that the spin states are stabilized (arrow-up) or unstabilized (arrow-down) due to the exchange and correlation interaction.

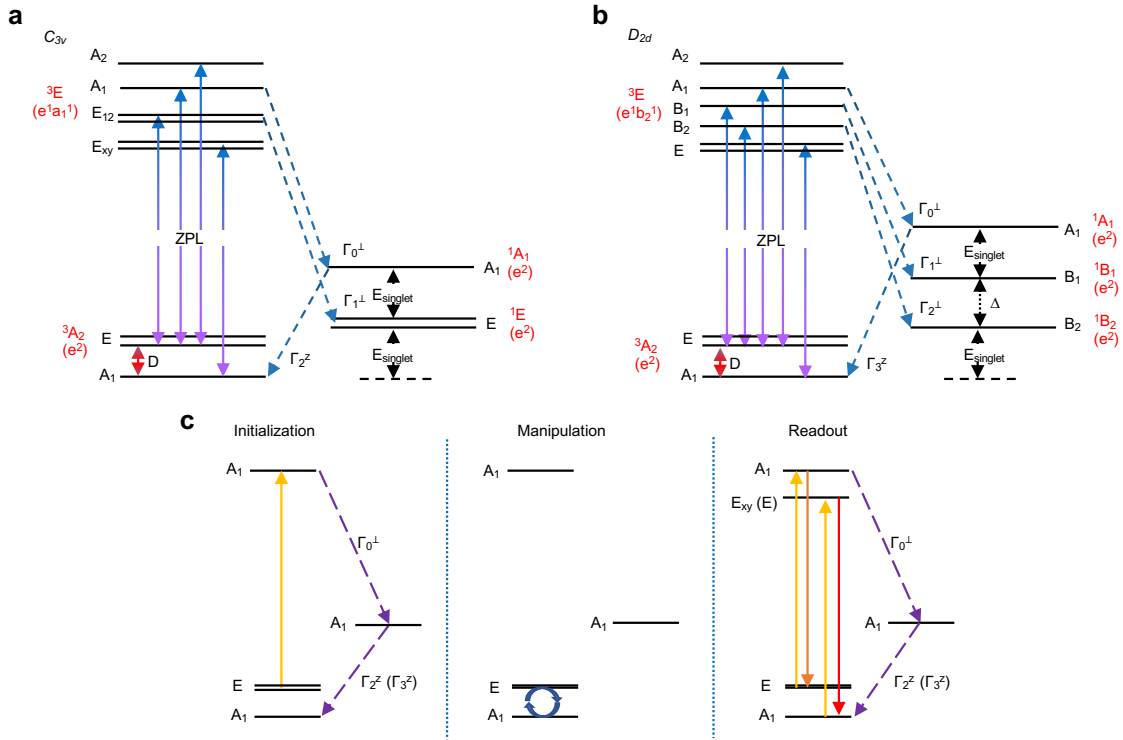

Figure 6 **Energy diagrams and qubit operation loop of defect qubit candidates with $C_{3v}$ and $D_{2d}$.** **a** Energy diagram of antisites with symmetry $C_{3v}$. The sublevels for the triplet states and the singlet states are denoted by IRs in $C_{3v}$. The allowed radiative transitions are denoted as the colored solid arrows, and the allowed irradiative intersystem crossings are denoted as the dashed arrows. The ZPL, $D$, and $E_{singlet}$ refer to the zero-phonon line, the ZFS $D$, and the energy difference bewteen $^3A_2$ and $^1E$. **b** Energy diagram of antisites with symmetry $D_{2d}$. The energy difference between $^1B_2$ and $^1B_1$ denoted as $\Delta$ is not determined here. **c** Qubit operational loop has three steps, initialization, manipulation, and readout. The absorption radiative processes are denoted as the yellow lines. In the readout, the emission (the red line) has a higher intensity than the emission (the orange line) due to the existence of intersystem crossings (the dashed purple lines) that weakens the intensity of the radiative transition.

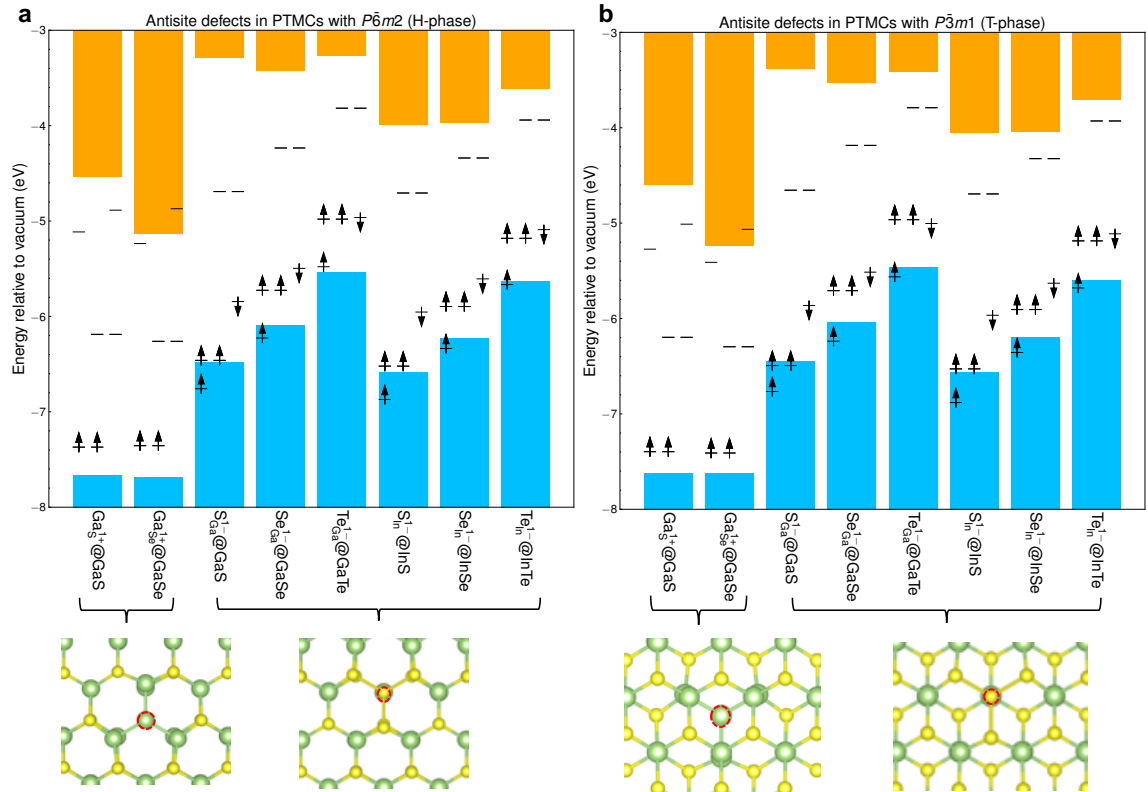

Figure 7 **Defect levels and structures of antisite defects in PTMCs. a** Defect levels and structures of $M_X^{1+}$ (M: Ga; X: S, Se) and $X_M^{1-}$ (X: S, Se, Te; M: Ga, In) in H-phase MX (M: Ga, In; X: S, Se, Te) with *P-6m2*. The upward/downward black arrows denote spin-up/spin-down electrons. Blue and orange bars denote valence and conduction bands, respectively. The red-circled atom denotes the location of antisite. **b** Defect levels and structures of $M_X^{1+}$ (M: Ga; X: S, Se) and $X_M^{1-}$ (X: S, Se, Te; M: Ga, In) in T-phase MX (M: Ga, In; X: S, Se, Te) with *P-3m1*.

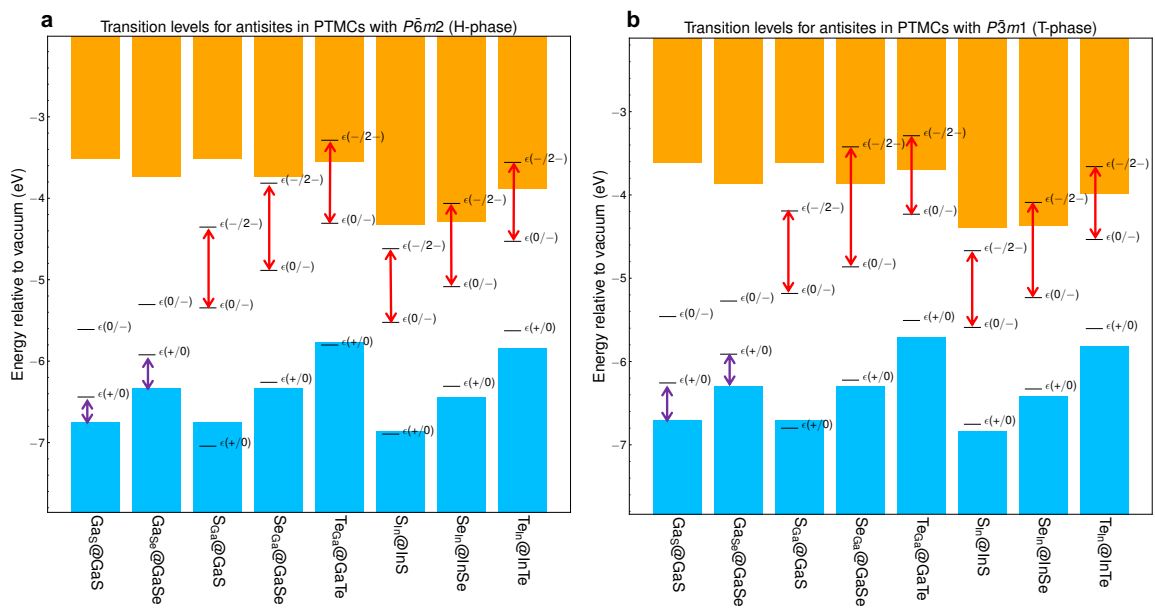

Figure 8 **Transition levels of antisite defects in PTMCs. a** Transition levels of $M_X$ (M: Ga; X: S, Se) and $X_M$ (X: S, Se, Te; M: Ga, In) in H-phase MX (M: Ga, In; X: S, Se, Te) with *P-6m2*. $\varepsilon(q/q')$ denotes the transition level from the charge state *q* to *q'*. The windows of Fermi levels that give rise to the stable charge states 1+ in the anion antisites and 1- in the cation antisites are the differences between VBM and $\varepsilon(+/0)$ (the purple double-arrow lines) and $\varepsilon(0/-)$ and $\varepsilon(-/2-)$ ) (the red double-arrow lines), respectively. **b** Transition levels of $M_X$ (M: Ga; X: S, Se) and $X_M$ (X: S, Se, Te; M: Ga, In) in T-phase MX (M: Ga, In; X: S, Se, Te) with *P-3m1*.

Table 1 **A full list of 471 defects in binary 2D materials with triplet ground states calculated using the r²SCAN functional.** The full spreadsheet for defects with triplet ground states calculated by r²SCAN is available in the Figshare repository https://figshare.com/s/876472cbc5ed8edf0990. The descriptions of the columns are shown in Supplementary Table 1.

Table 2 **Statistics of defect types and charge states for triplet defects calculated using the r²SCAN functional.**

| Defect type | Charge state | Number of entries | Fraction (%) |
| --- | --- | --- | --- |
| Antisite | 1- | 123 | 26 |
| | 0 | 77 | 16 |
| | 1 | 77 | 16 |
| Vacancy | 1- | 59 | 13 |
| | 0 | 90 | 19 |
| | 1 | 45 | 10 |

Table 3 **Statistics of defect types and charge states for triplet defects calculated by r2SCAN with ELUDL-HODL larger than or equal to 0.5 eV.**

| Defect type | Charge state | Number of entries | Fraction (%) |
| --- | --- | --- | --- |
| Antisite | -1 | 59 | 39 |
| | 0 | 47 | 31 |
| | 1 | 30 | 20 |
| Vacancy | -1 | 7 | 5 |
| | 0 | 3 | 2 |
| | 1 | 6 | 4 |

Table 4 **A full list of 141 defects with triplet ground states calculated using HSE06 functional.** The full spreadsheet for defects with triplet ground states calculated by HSE06 functional is available in the Figshare repository https://figshare.com/s/876472cbc5ed8edf0990. The columns and their descriptions are shown in Supplementary Table 2.

Table 5 **The full list of 45 quantum defect candidates.** The full spreadsheet for quantum defect candidates is available in the Figshare repository https://figshare.com/s/876472cbc5ed8edf0990. The columns and their descriptions are shown in Supplementary Table 3.

Table 6 **Term symbols and spatial wavefunctions of defect symmetry groups $C_{3v}$ and $D_{2d}$.** $e_x$ and $e_y$ are two degenerate defect levels.

| Defect symmetry | $C_{3v}$ | $D_{2d}$ |
|---|---|---|
| Configuration | $e^2$ | $e^2$ |
| Triplet ground state | $^3A_2 = e_x e_y - e_y e_x|$ | $^3A_2 = e_x e_y - e_y e_x$ |
| Singlet states | $^1E_1 = e_x e_x - e_y e_y$ | $^1B_1 = e_x e_x - e_y e_y$ |
|  | $^1E_2 = e_x e_y + e_y e_x$ | $^1B_2 = e_x e_y + e_y e_x$ |
|  | $^1A_1 = e_x e_x + e_y e_y$ | $^1A_1 = e_x e_x + e_y e_y$ |
| No. of candidates | 34 | 11 |

Table 7 **Sublevels and intersystem crossings between triplet states and singlet states.** The table shows the sublevels in the triplet excited state ($T^*$), in the singlet states ($S$), and in the triplet ground state ($T$) of a defect imposing $C_{3v}$ or $D_{2d}$. The intersystem crossings are labeled by $\{\Gamma_0^\perp, \Gamma_1^\perp, \Gamma_2^z\}$ for the antisites with $C_{3v}$ and $\{\Gamma_0^\perp, \Gamma_1^\perp, \Gamma_2^\perp, \Gamma_3^z\}$ for the antisites with $D_{2d}$. Those intersystem crossings with the perpendicular symbol are mediated by $H_{so}^\perp$, and those with the parallel symbol are mediated by $H_{so}^{//}$.

| Defect symmetry | $C_{3v}$ | $D_{2d}$ |
|---|---|---|
| Sublevels ($T^*$) | $^3E: (A_2, A_1, E_{12}, E_{xy})$ | $^3E: (A_2, A_1, B_1, B_2, E)$ |
| Sublevels ($S$) | $^1E_1: E$ | $^1B_1: B_1$ |
|  | $^1E_2: E$ | $^1B_2: B_2$ |
|  | $^1A_1: A_1$ | $^1A_1: A_1$ |
| Sublevels ($T$) | $^3A_2: (E, A_1)$ | $^3A_2: (E, A_1)$ |
| Intersystem crossings | $\Gamma_0^\perp: (A_1)\ T^* \to S$ | $\Gamma_0^\perp: (A_1)\ T^* \to S$ |
|  | $\Gamma_1^\perp: (E_{12})\ T^* \to S\ (E)$ | $\Gamma_1^\perp: (B_1)\ T^* \to S$ |
|  | $\Gamma_2^z: (A_1)\ S \to T$ | $\Gamma_2^\perp: (B_2)\ T^* \to S$ |
|  |  | $\Gamma_3^z: (A_1)\ S \to T$ |

Table 8 **Parameters of the electronic structures for the pristine and defective PTMCs.** The band gaps of the pristine materials are calculated using HSE06 functional. $E_{triplet-singlet}$ denotes the total energy difference between the spin singlet and triplet of the antisites. $E_F$ window denotes the Fermi-level window that gives rise to the charge states 1+ or 1-.

| Pristine PTMCs | Bandgap (eV) | Defect name | Charge | $E_{triplet-singlet}$ | $E_F$ window |
|---|---|---|---|---|---|
| GaS (H) | 3.222 | $Ga_S^{1+}$ | 1 | -0.157 | 0.31 |
|  |  | $S_{Ga}^{1-}$ | -1 | -0.184 | 0.99 |
| GaS (T) | 3.104 | $Ga_S^{1+}$ | 1 | -0.158 | 0.45 |
|  |  | $S_{Ga}^{1-}$ | -1 | -0.191 | 0.99 |
| GaSe (H) | 2.641 | $Ga_{Se}^{1+}$ | 1 | -0.036 | 0.41 |
|  |  | $Se_{Ga}^{1-}$ | -1 | -0.177 | 1.07 |
| GaSe (T) | 2.491 | $Ga_{Se}^{1+}$ | 1 | -0.149 | 0.38 |
|  |  | $Se_{Ga}^{1-}$ | -1 | -0.18 | 1.44 |
| GaTe (H) | 2.11 | $Te_{Ga}^{1-}$ | -1 | -0.147 | 1.02 |
| GaTe (T) | 2.058 | $Te_{Ga}^{1-}$ | -1 | -0.149 | 0.94 |
| InS (H) | 2.553 | $S_{In}^{1-}$ | -1 | -0.191 | 0.90 |
| InS (T) | 2.482 | $S_{In}^{1-}$ | -1 | -0.194 | 0.92 |
| InSe (H) | 2.204 | $Se_{In}^{1-}$ | -1 | -0.183 | 1.02 |
| InSe (T) | 2.105 | $Se_{In}^{1-}$ | -1 | -0.186 | 1.14 |
| InTe (H) | 2.014 | $Te_{In}^{1-}$ | -1 | -0.155 | 0.97 |
| InTe (T) | 1.909 | $Te_{In}^{1-}$ | -1 | -0.158 | 0.88 |